# Freestanding Antiferromagnetic Oxide Membranes: Synthesis and Characterization of $Cr_2O_3$

*Ella Blake[1,2], Tiffany C. Wang[2,3], Yi Cui[2,4], Minyong Han[2], Yinchuan Lv[2], Christopher T. Parzyck[2], Octave Duros[2], Yan Che[2], Matthias C. Hoffmann[5], Wei-Sheng Lee[2], Yi Cui[2,4], Harold Y. Hwang[2,3]*

[1]Department of Physics, Stanford University; Stanford, CA 94305, United States.

[2]Stanford Institute for Materials and Energy Sciences, SLAC National Accelerator Laboratory; Menlo Park, CA 94025, United States.

[3]Department of Applied Physics, Stanford University; Stanford, CA 94305, United States.

[4]Department of Materials Science and Engineering, Stanford University; Stanford, CA 94305, United States.

[5]Linac Coherent Light Source, SLAC National Accelerator Laboratory; Menlo Park, CA 94025, United States.

**ABSTRACT**

Antiferromagnetic (AFM) insulators stand out as a promising class of materials for fast switching, energy-efficient magnon-based technology, but our understanding of how to tune these properties with the lattice remains limited. As an AFM insulator with strong magnetoelectric response, $Cr_2O_3$ membranes stand out as a promising platform to explore multi-modal tunability via strain. However, fabricating these membranes is challenging due to the scarcity of etchable sacrificial layers with compatible lattice constants and sufficient surface quality. We address this issue by utilizing $La_{0.7}Sr_{0.3}MnO_3$ (LSMO) as a sacrificial layer, enabling successful fabrication of millimeter-scale $Cr_2O_3$ membranes. Substrate-free characterization reveals that strain due to the lattice mismatch during growth is released by the formation of small polycrystalline domains, preserving single-crystalline order over ~90% of the membrane area. Bulk-like structural properties are confirmed by transmission electron microscopy, X-ray diffraction, Raman spectroscopy, and second-harmonic generation spectroscopy. Platinum Hall measurements reveal an above-room-temperature Néel transition. These results establish $Cr_2O_3$ membranes as a viable platform for strain-tuning antiferromagnetism and magnetoelectricity.

## INTRODUCTION

Antiferromagnetic (AFM) crystals have garnered attention recently in the field of magnonics—or magnon-based devices—due to their robustness against external magnetic fields and hosting high frequency magnon modes relative to ferromagnets (FM)[1–8]. These AFM magnon modes often lie in the hundreds of GHz to THz range, with correspondingly higher switching speed limits than those of FMs[1,3,4]. AFM insulators also facilitate the propagation of magnons without charge carrier transport, alleviating the energy cost due to Joule heating. Thus, AFM insulator-based devices stand to be both faster and more energy efficient than their FM counterparts[1,2,8,9].

Strain has emerged as a robust means of controlling the electronic and magnetic properties of complex oxide membranes via their coupling to the lattice. This avenue of tunability has proven useful in a variety of oxide systems to explore correlated electron behavior, including superconductivity[10–12]. In the case of $Cr_2O_3$, magnetoelectricity sets the backdrop for a rich interplay between magnetic, electronic, and lattice degrees of freedom[9,13–17]. The challenge, however, lies in achieving controllable and scalable strain while preserving high crystalline quality. For thin films, most relevant for device integration, epitaxial strain has been used to tune these properties[18–20]. This method is limited by the degree of strain imparted by the substrate, and care must be taken to isolate strain-induced effects from those induced by disorder or thickness[19,20]. To address this, we have developed the synthesis of $Cr_2O_3$ as a flexible membrane: a thin film free from its substrate. Due to the mechanical flexibility of this format, we anticipate a greater and continuous range of strain can be applied to control the properties of this AFM insulator[10,21]. Furthermore, freestanding membranes facilitate substrate-free characterization. Fabricating $Cr_2O_3$ in this form, however, is complicated by the extreme lattice mismatch with many conventional etchable sacrificial layers[10,11]. Moreover, many existing oxide membrane synthesis methods employ an $SrTiO_3$ capping layer, which is undesirable when studying a dielectric membrane due to the strong dielectric contributions from the capping layer[10,11,22]. In this work, we successfully stabilize $Cr_2O_3$ membranes without the requirement of an $SrTiO_3$ cap. We characterize their structural and magnetic properties using transmission electron microscopy (TEM), X-ray diffraction (XRD), Raman spectroscopy, second-harmonic generation (SHG) spectroscopy, and Pt

Hall transport, establishing $Cr_2O_3$ membranes as a promising first step towards flexible AFM insulator-based magnonic devices.

## EXPERIMENTAL METHODS

### Synthesis of $Cr_2O_3$

Here we describe a method of synthesizing millimeter-scale crack-free membranes of $Cr_2O_3$ using an acid-soluble sacrificial layer. The process flow of transferring the membrane to a new substrate is visualized in Fig. 1a. Generally, to achieve such large-scale crack-free regions, minimization of the lattice mismatch between each layer of the heterostructure is desirable. Despite a severe (>10%, see Table 1) mismatch with the sacrificial layer, we successfully stabilized crack-free freestanding *c*-plane $Cr_2O_3$ (schematic crystal structure shown in Fig. 1b) in membrane form enabled by a unique strain mitigation mechanism explored in the results section.

We selected $La_{0.7}Sr_{0.3}MnO_3$ (LSMO), a distorted perovskite, as the etchable sacrificial layer[23] due to its ability to grow with low surface root-mean-square (rms) roughness (0.553 nm) in the (111) pseudo-cubic direction, providing a smooth interface for subsequent $Cr_2O_3$ epitaxy. The hexagonal structure of $Cr_2O_3$ is shown in Fig 1b. A ~20 nm layer was grown directly on an $SrTiO_3$(111) substrate via pulsed laser deposition at a temperature of 700 °C under oxygen partial pressure of $P_{O_2} = 100$ mTorr. In the same synthesis chamber, we then grew ~25 nm of $Cr_2O_3$ onto the sacrificial layer at 600 °C under $P_{O_2} = 20$ mTorr. For comparison, we additionally synthesized epitaxial $Cr_2O_3$ thin films, grown directly on *c*-plane $Al_2O_3$ substrates. This deposition occurred at a temperature of 700 °C under $P_{O_2} =$ 10 mTorr.

After growth, ~1.2 μm polymethyl methacrylate (PMMA) was spin-coated on top of the $Cr_2O_3$/LSMO heterostructure as a supportive layer during the etching process. The heterostructure was immersed in a dilute potassium iodide/ hydrochloric acid solution (0.6 M KI, 0.06 M HCl) for two days, selectively dissolving the LSMO and releasing the membrane from the substrate. The resulting membrane was then transferred onto either an $SiO_2$/Si(001) wafer or a silicon nitride window for measurement. To increase adhesion with the new substrate surface, the membrane and substrate were heated on a hot plate at 70 °C for 10 minutes and then 110 °C for an hour. Next, the PMMA supporting layer was carefully rinsed off with acetone and isopropyl alcohol (IPA), revealing the bare, continuous $Cr_2O_3$ membrane underneath, as seen in Fig. 1c. Given the aforementioned severe lattice mismatch through the membrane heterostructure, we further confirmed the lack of microscopic cracking via atomic force microscopy. In Fig. 1d, we see a continuous membrane placed onto an $SiO_2$/Si(001) wafer with no signs of breaks in the membrane.

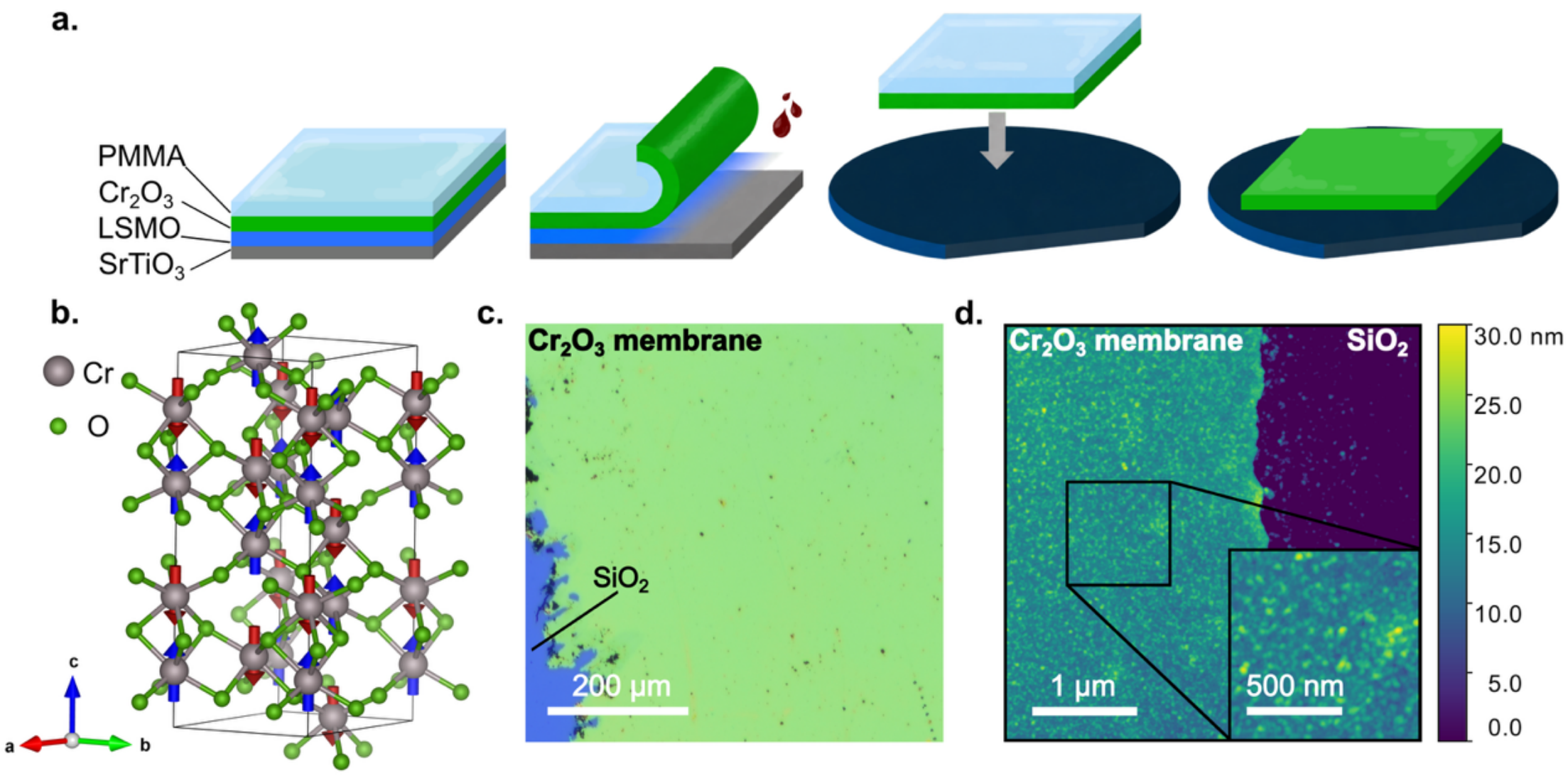

FIG. 1 Synthesis of freestanding $Cr_2O_3$ membranes. (a) Schematic of membrane transfer process. The sacrificial LSMO layer is chemically etched away to enable transfer onto various other substrates. (b) Structure of *c*-plane $Cr_2O_3$. Antiferromagnetic ordering shown by the orientation of blue and red arrows. The Néel vector points along the hexagonal *c*-axis. (c) Optical microscope image of membrane transferred onto $SiO_2$/Si(001). (d) Atomic force microscope image of a membrane transferred onto $SiO_2$/Si(001). Inset shows the surface over a narrower field of view.

## Polarization-resolved SHG spectroscopy

SHG measurements were performed using an 800-nm Ti:sapphire laser with a repetition rate of 66 MHz and a pulse duration of approximately 50 fs. The fundamental beam was focused onto the sample using a 20× objective, producing a spot size of approximately 2 μm. The reflected SHG signal at 400 nm was collected in an epi-detection geometry using the same objective and detected with a Hamamatsu photomultiplier tube. The polarization of the incident fundamental beam was controlled using a half-wave plate, while a linear polarizer in the detection path was used to select the polarization of the SHG signal. For polarization-resolved measurements, the fundamental and SHG polarizations were maintained parallel to each other and rotated together with respect to the crystallographic axes of the sample. All SHG measurements were performed at 295 K.

### Platinum Hall Device Fabrication

We made use of the strong spin-orbit coupling of platinum to detect Néel transition at the surface of the $Cr_2O_3$[20,24–26] by fabricating Pt devices onto the surfaces of the transferred membranes. Specifically, the Hall magnetoresistance of the Pt is sensitive to the magnetism at the Pt/$Cr_2O_3$ interface, so we used thin (5 nm) Pt to reduce the relative contribution of bulk transport through the Pt. After the membrane was transferred onto the $SiO_2$ and cleaned, platinum was deposited via evaporation over the entire surface before patterning 5 μm Hall crosses with AZ1512 photoresist. This method protects the sensitive Pt/$Cr_2O_3$ interface from contamination. The device pattern was then etched into the platinum via ion milling. Next, the device was annealed at 400 °C for 30 minutes at ambient pressure in a tube furnace, inducing crystallization of the platinum in the (111) orientation.

## RESULTS & DISCUSSION

### Overcoming heterostructure lattice mismatch

$Cr_2O_3$ films are conventionally grown on sapphire $Al_2O_3$ substrates due to their isostructural relationship and relatively small (−3.9%) lattice mismatch[27,28]. However, few candidates for sacrificial layers can be both grown on sapphire and selectively etched without damaging the $Cr_2O_3$. For our membranes, we elected to grow the membrane heterostructure on $SrTiO_3$(111) substrates with an effective lattice mismatch of 11.48%, as seen in Table 1. The sacrificial LSMO layer is more closely matched to the substrate, and when grown at thickness of 20 nm, it conforms to the substrate lattice. This means that the $Cr_2O_3$ layer is under significant epitaxial strain; nevertheless, we successfully synthesized large-scale continuous $Cr_2O_3$ membranes.

TABLE 1. Effective in-plane lattice parameters of substrates and sacrificial layer used in this work.

| Material | Lattice Constant (Å) | Mismatch with $Cr_2O_3$ (%) | Ref. |
|---|---|---|---|
| $Cr_2O_3$ | 4.953 | -- | [27] |
| $Al_2O_3$ | 4.759 | −3.9 | [28] |
| $SrTiO_3$ | $a \cdot \sqrt{2} = 5.522$ | 11.48 | [29] |
| $La_{0.7}Sr_{0.3}MnO_3$ | $a \cdot \sqrt{2} = 5.487$ | 10.78 | [30] |

We use planar-view atomic resolution TEM to shed light on the microscopic mechanism of strain mitigation in the Cr2O3 membranes. The membrane was placed onto a 200 nm thick Norcada silicon nitride window with 5 μm pores, as seen in Fig. 2a. At higher magnification, Fig. 2b, we observed neatly ordered rows of atoms across most of the membrane, but there are also streak-like patches where the crystalline order breaks down; these disordered regions are highlighted in green. This field of view is representative of the typical macroscopic defect density across the membrane. The projected area of these regions represents 10.2% of the total area in this field of view. The inset shows a closer view of the ordered region. The disordered regions appear to facilitate the release of epitaxial strain rather than simply demarcating a region of disorder between domains. This interpretation is supported by the observation that the alignment of rows of atoms in the ordered regions on either side of a defect is identical, inconsistent with domain boundary formation. Figures 2c and 2d show the selected area diffraction patterns taken in the ordered and disordered regions, respectively. The diffraction pattern from the ordered region shows sharp spots consistent with the hexagonal $Cr_2O_3$ structure. In contrast, the pattern from the disordered region shows a superposition of rotationally offset hexagonal patterns, indicating local polycrystallinity. These disordered regions enable the majority of the $Cr_2O_3$ membrane to remain single-crystalline despite the extreme epitaxial strain imposed during growth.

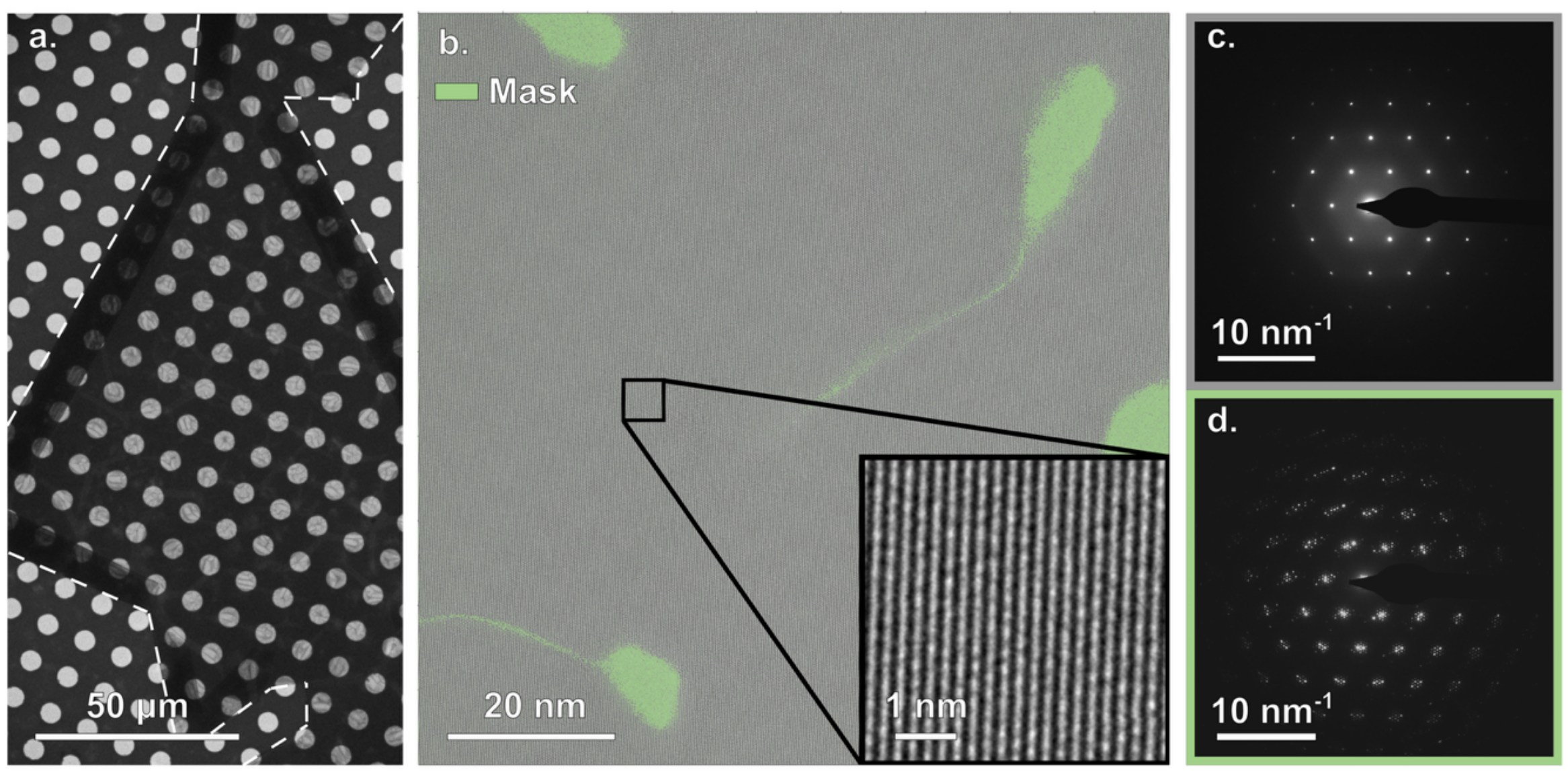

FIG. 2 Transmission electron microscopy images of a freestanding $Cr_2O_3$ membrane. (a) Wide view image showing the membrane (outlined with white dashed line) on top of a porous silicon nitride window. (b) Atomic resolution image of a membrane. Inset shows a narrow view. Disordered regions are highlighted in green (c) Selected area diffraction pattern taken from an ordered region in (b). (d) Selected area diffraction pattern taken from a disordered region in (b).

## Robust structural and magnetic properties of membranes

To demonstrate these $Cr_2O_3$ membranes constitute a viable platform for studying the intrinsic physics in the material, we employed a XRD, Raman spectroscopy, and SHG spectroscopy (Fig. 3) to establish structural quality. We use Pt Hall transport (Fig. 4) to confirm magnetic quality. Examining the X-ray diffraction (XRD) pattern from the membrane heterostructure before sacrificial layer etching, we identified the three peaks corresponding to the substrate, LSMO sacrificial layer, and $Cr_2O_3$ in Fig. 3a. Here the *c*-axis reflection of the $Cr_2O_3$ is shifted in response to the in-plane strain from the sacrificial layer and substrate below. We collected XRD data from the released membrane in Fig. 3a. Since the membrane no longer has an epitaxial relationship with the substrate, its lattice is unpinned and relaxes back to its bulk position, as seen in the peak shift towards the bulk value.

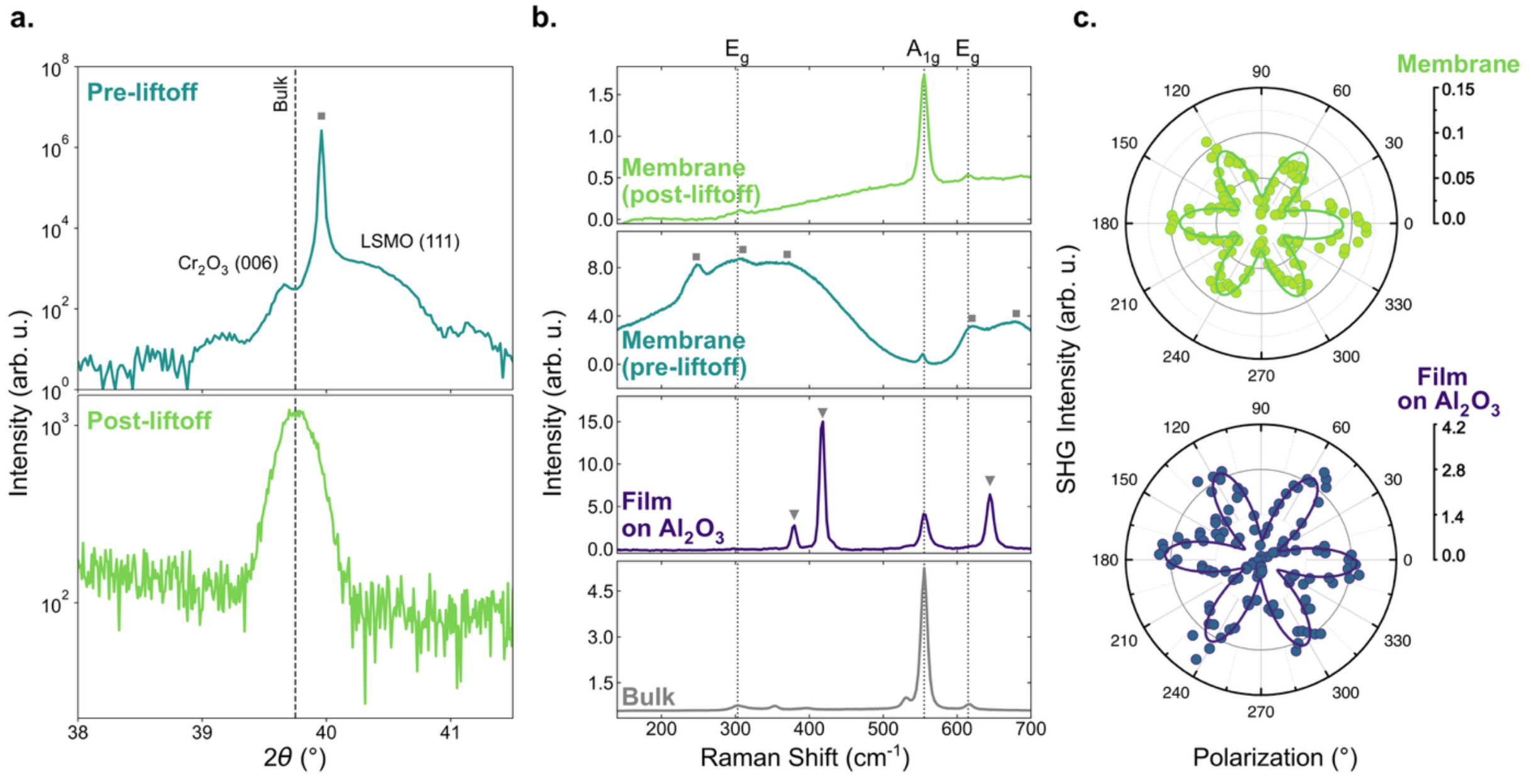

FIG. 3 X-ray and optical characterization of membranes. (a) XRD $\theta-2\theta$ symmetric scan of $Cr_2O_3$ for membrane prior to etching the sacrificial layer (top) and from the lifted-off membrane (bottom). $SrTiO_3$(111) substrate reflection is denoted by a square. Bulk $Cr_2O_3$ *c*-axis peak position is marked by dashed vertical line. (b) Raman spectroscopy of $Cr_2O_3$ as a membrane (green), an as-grown heterostructure with the LSMO sacrificial layer (teal), an epitaxial film on $Al_2O_3$ (purple), and a bulk crystal (gray). Spectra have been normalized to $Cr_2O_3$ thickness and laser power. $Cr_2O_3$ $A_{1g}$ and $E_g$ modes are marked with black vertical dotted lines. Substrate modes from $SrTiO_3$ and $Al_2O_3$ are marked with squares and triangles, respectively. (c) SHG characterization of a $Cr_2O_3$ film and membrane. Polarization dependence of the SHG intensity measured from a 20-nm-thick membrane transferred onto a *c*-sapphire substrate (top) and from an epitaxial

100-nm-thick film grown on a *c*-sapphire substrate (bottom). An angle of 0° corresponds to a polarization parallel to the crystallographic *a*-axis.

To probe lattice dynamics in the membranes, we use Raman spectroscopy to characterize the phonon modes of $Cr_2O_3$[31–33]. The $A_{1g}$ mode corresponds to a symmetry-preserving rotational displacement of the oxygen atoms in the unit cell[31]. We focus on this mode as it is the most prominent in bulk $Cr_2O_3$ and because shifts in its position serve as a sensitive indicator of strain. In Fig. 3b, we compare a thin film grown on sapphire, with a lower degree of epitaxial strain, to the $Cr_2O_3$ in the as-grown heterostructure (pre-liftoff) and a released free-standing membrane (post-liftoff). The additional peaks in the thin film spectrum correspond to modes present in the sapphire substrate; similarly, the broad features in the pre-liftoff spectrum correspond to the $SrTiO_3$ substrate. There is a 2 $cm^{-1}$ redshift in the $A_{1g}$ peak for the pre-liftoff membrane sample, consistent with strain imposed by the substrate. Across the thin film, bulk, and post-liftoff membrane samples, the $A_{1g}$ peak shows minimal shift, confirming the robustness of the membrane's structural properties as well as the relaxed strain state. The $E_g$ modes visible in both the bulk and membrane spectra correspond to symmetry-lowering displacements of the oxygen ions about a central chromium ion[34]. Notably, these less-prominent $E_g$ peaks cannot be discerned in the thin films but are visible in the membrane. These results underscore the utility of membrane fabrication towards substrate-free characterization while remaining in the thin regime.

The sixfold rotational symmetry observed in the SHG linear polarization dependence of $Cr_2O_3$ has been attributed to the simultaneous contributions of electric and magnetic dipole transitions[35]. To confirm that this symmetry is preserved in membrane form, we performed polarization-resolved SHG microscopy on both the transferred membrane and a reference $Cr_2O_3$ thin film, both on *c*-plane $Al_2O_3$[36,37]. This substrate is ideal for both the thin film and transferred membrane sample due to its minimal contribution to the measured SHG signal. Polarization-resolved SHG provides a complementary probe of the crystallographic integrity of the transferred membrane. As shown in Fig. 3c, both the as-grown thin film and the membrane exhibit a sixfold polarization dependence, demonstrating that the characteristic nonlinear optical anisotropy of $Cr_2O_3$ is preserved after membrane transfer[35,38]. We speculate that slight distortions in the sixfold symmetry of the SHG polarization-dependence result from minor local strain imposed when placing the membrane onto the $Al_2O_3$ substrate. The combination of XRD, and small spot-size optical probes (Raman, SHG) show that the membrane retains crystallographic uniformity and nonlinear optical characteristics comparable to those of the epitaxial thin film over a range of length scales.

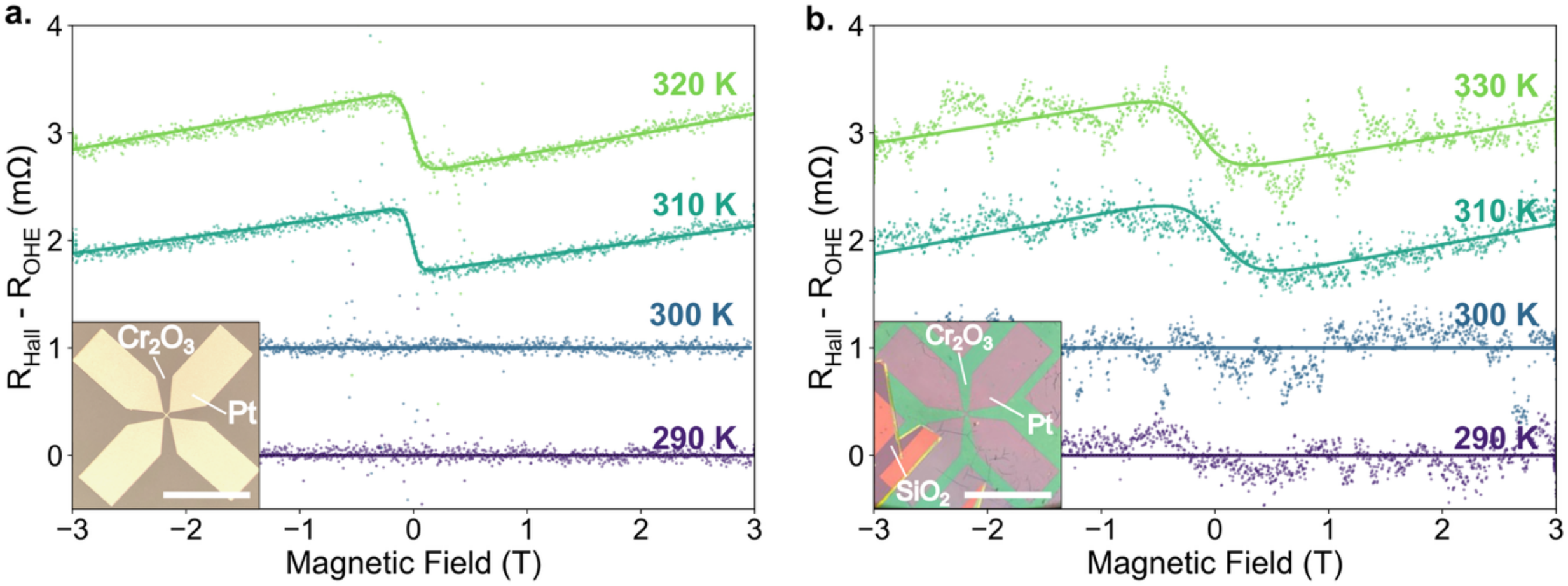

FIG. 4 Hall measurements with Pt (5nm)/$Cr_2O_3$ (25 nm) heavy metal-antiferromagnet interface devices patterned onto a thin film (a) and a membrane transferred onto a $SiO_2$/Si wafer (b) at several temperatures. Magnetic field is applied out-of-plane relative to the sample. Hall resistance is plotted with the linear ordinary

Hall contribution, $R_{OHE}$, subtracted[25]. The characteristic step in Pt Hall resistance disappears at the $Cr_2O_3$ Néel temperature. All data are vertically offset for visual clarity. Insets show optical images of 5 μm-wide Pt cross devices on thin film (a) and on a membrane (b); scale bar is 200 μm.

Much of the interest surrounding $Cr_2O_3$ centers on its magnetism as the canonical antiferromagnetic magnetoelectric, so it is important to demonstrate that through the membrane synthesis process we have preserved the near-room-temperature Néel transition ( $T_N^{Bulk} \approx 307$ K )[13–15]. To detect the onset of antiferromagnetic ordering, we utilized platinum Hall devices. When platinum is measured on a non-magnetically-ordered surface, the Hall magnetoresistance follows a characteristic step-like field dependence due to the Hanle magnetoresistance (HMR) contribution[20,25]. HMR arises when an applied charge current generates a transverse pure spin current via the spin Hall effect (SHE), producing a non-equilibrium spin accumulation at the film surfaces with polarization perpendicular to both the charge and spin current directions[39]. When a magnetic field is applied perpendicular to this spin polarization, the accumulated spins undergo Larmor precession and subsequent dephasing[39]. The reoriented spin accumulation is then converted into a transverse charge voltage via the inverse spin Hall effect (ISHE), resulting in a field-dependent Hall resistance[25]. This serves as a baseline against which modifications due to antiferromagnetic ordering at the interface can be identified. Below $T_N$, the uncompensated moments at the top surface of the antiferromagnetically-ordered $Cr_2O_3$ serve to pin the spin accumulation direction via an exchange coupling. This effectively suppresses the HMR in the Pt once Néel order has been established[20,24,25]. Therefore, the temperature at which this suppression occurs corresponds directly to the Néel transition temperature of the $Cr_2O_3$ at the interface[20,24]. The sensitivity of this measurement to the heavy metal-antiferromagnet interface makes this measurement technique exceptionally useful for diagnosing magnetic transitions in antiferromagnets, where Néel order is otherwise challenging to observe[4,20,24].

We measured platinum Hall devices on $Cr_2O_3$ in thin film and membrane form (Figs. 4a and 4b, respectively), with the linear ordinary Hall effect contribution subtracted. Optical microscope images of both devices are included in the insets. The HMR in the platinum is observed in both film and membrane form at and above 310 K and appears fully suppressed at 300 K, indicating a Néel transition consistent with bulk values[13,15]. The similar signal amplitude above $T_N$ and the complete suppression down to the noise floor in both film and membrane data is consistent with a similarly sharp transition in each form. The preservation of a sharp, bulk-like $T_N$ demonstrates the compatibility of $Cr_2O_3$ membranes with room-temperature magnonic-devices.

**CONCLUSIONS**

Here we present the synthesis and characterization of $Cr_2O_3$ membranes, including the fabrication of micro-transport devices used to observe the Néel transition. Utilizing an acid-etchable sacrificial $La_{0.7}Sr_{0.3}MnO_3$ layer, we stabilized these membranes with long-range crystalline order despite a large lattice mismatch. We find that the strain resulting from this mismatch is alleviated through the formation of small polycrystalline regions, which preserve the single-crystalline integrity of the surrounding ~90% of the membrane. X-ray diffraction, Raman spectroscopy, SHG polarization-dependence, and platinum Hall measurements all confirm bulk-like structural, optical, and magnetic properties, including a Néel transition between 300 and 310 K. The demonstrated properties of these membranes indicate their promise as a flexible platform for studying antiferromagnetism and magnetoelectricity under strain, as well as integration with other materials families.

## AUTHOR INFORMATION

### Conflict of Interest

The authors have no conflicts to disclose.

### Corresponding Author

Ella Blake (erblake@stanford.edu); Harold Y. Hwang (hyhwang@stanford.edu)

### Author Contribution

**Ella Blake:** Conceptualization (equal); Data curation (lead); Visualization (lead); Writing (lead); $La_{0.7}Sr_{0.3}MnO_3$ target synthesis (lead); Raman spectroscopy (equal); X-ray diffractometry (lead); Pt device fabrication (lead); Transport measurements (lead). **Tiffany C. Wang**: Conceptualization (equal); Thin film and membrane synthesis (supporting); Writing – review & editing (equal). **Yi Cui**: TEM data collection (lead) **Minyong Han**: Conceptualization (equal); $Cr_2O_3$ target synthesis (lead); Thin film and membrane synthesis (supporting). **Yinchuan Lv**: SHG measurements (lead); Writing – review & editing (supporting). **Christopher T. Parzyck**: Raman spectroscopy (lead). **Octave Duros**: Raman spectroscopy (supporting). **Yan Che**: Atomic force microscopy (supporting). **Matthias C. Hoffmann**: SHG measurements (supporting). **Wei-Sheng Lee**: Raman spectroscopy (supporting). **Yi Cui**: TEM data collection (supporting). **Harold Y. Hwang**: Conceptualization (equal); Funding acquisition (lead); Writing – review & editing (equal).

## ACKNOWLEDGEMENTS

We appreciate the fruitful discussions with Yonghun Lee and Arturas Vailionis. This work was supported by the U.S. Department of Energy, Office of Science, Basic Energy Sciences at SLAC National Laboratory under Contract No. DE-AC02-76SF00515, as part of the Center for Energy Efficient Magnonics, an Energy Frontier Research Center, and other core programs.

## DATA AVAILABILITY

The data that support the findings of this study are available from the corresponding author upon reasonable request.